\documentclass{article}
\usepackage{spconf,amsmath,graphicx,hyperref,cite}
\usepackage{ieeetrantools}
\usepackage{etoolbox}
\apptocmd{\thebibliography}{\setlength{\itemsep}{0pt}}{}{}

\title{ICASSP 2026 URGENT Speech Enhancement Challenge}
\name{
\begin{tabular}{c}
Chenda Li$^{1}$, Wei Wang$^{1}$, Marvin Sach$^{2}$, Wangyou Zhang$^{1}$, Kohei Saijo$^{3}$, Samuele Cornell$^{4}$, \\
Yihui Fu$^{2}$, Zhaoheng Ni$^{5}$, Tim Fingscheidt$^{2}$, Shinji Watanabe$^{4}$, Yanmin Qian$^{1}$\thanks{Thanks to Robin Scheibler and Anurag Kumar for their contributions to the challenge organization.}
\end{tabular}
}
\address{$^{1}$Shanghai Jiao Tong University, China; $^{2}$Technische Universität Braunschweig, Germany; \\
$^{3}$Waseda University, Japan;
$^{4}$Carnegie Mellon University, USA; $^{5}$Meta, USA}
\begin{document}
\bstctlcite{IEEEexample:BSTcontrol} 

\ninept
\maketitle
\begin{abstract}
The ICASSP 2026 URGENT Challenge advances the series by focusing on universal speech enhancement (SE) systems that handle diverse distortions, domains, and input conditions. This overview paper details the challenge's motivation, task definitions, datasets, baseline systems, evaluation protocols, and results. The challenge is divided into two complementary tracks. Track 1 focuses on universal speech enhancement, while Track 2 introduces speech quality assessment for enhanced speech. The challenge attracted over 80 team registrations, with 29 submitting valid entries, demonstrating significant community interest in robust SE technologies.
\end{abstract}
\vspace{-5pt}
\section{Introduction}
\label{sec:intro}
\vspace{-5pt}

Speech enhancement (SE) aims at improving the intelligibility and quality of speech degraded by noise, reverberation, and other distortions \cite{loizouSpeechEnhancementTheory2007}. SE has advanced significantly with neural networks \cite{wangSupervisedSpeechSeparation2018a}, yet most prior work evaluates under matched conditions and limited distortions, leading to systems that struggle with unseen acoustic environments, diverse speaker characteristics, or unseen distortion types. This lack of generalizability hinders real-world deployment, where speech varies widely in content, languages, and styles.
To address these challenges, the URGENT Challenge (\textbf{U}niversal, \textbf{R}obust, and \textbf{G}eneralizable speech \textbf{E}nhanceme\textbf{NT}) \cite{zhangURGENTChallengeUniversality2024,koheiurgent2025} was launched to foster universal SE models that can adapt to diverse distortions, input formats, languages, etc. Building on previous editions that focused on robustness and multilinguality
\cite{zhangURGENTChallengeUniversality2024,koheiurgent2025}, the ICASSP 2026 URGENT Challenge emphasizes data curation and expanded speech diversity to overcome the limitations of current neural network approaches.

Compared with previous URGENT challenges, the key focus and contributions of this challenge include:
\textbf{1) Data Curation for Scalability:} Recent studies show diminishing returns from simply increasing dataset size without curation \cite{zhangPerformancePlateausComprehensive2024,gonzalezEffectTrainingDataset2024a,koheiurgent2025}. This challenge encourages the use of advanced data selection techniques to maximize performance gains \cite{liLessMoreData2025}. \textbf{2) Expanded Speech Diversity:} Beyond acoustics and languages, the challenge includes diversity of age, accents, whispered/singing voices, and emotional expressions, reflecting real-world variability.
\textbf{3) Perceptual Quality Assessment:} We newly introduce Track 2 to promote Mean Opinion Score (MOS) prediction for SE-processed speech, using large-scale human-rated datasets distinct from existing MOS prediction challenges\cite{uni-versa-ext}.

The challenge ran from September to November 2025, with evaluation phases including validation, non-blind test, and blind test. The challenge attracted 80 registered teams to Track 1 and 77 to Track 2, with 29 teams submitting valid entries across both tracks. Participants submitted systems via an online leaderboard. In the blind test of Track 1, the top-6 systems from the objective ranking proceed to subjective testing, where human listeners on the Amazon Mechanical Turk platform assess speech quality using the ITU-T P.808 Absolute Category Rating (ACR) and Comparison Category Rating (CCR) \cite{ITU-P808,P808-Sach2025}. MOS and Comparison Mean Opinion Scores (CMOS) on the blind test set determine the final rankings.

\vspace{-5pt}
\section{Challenge Description}
\vspace{-5pt}


\textbf{Track 1: Universal Speech Enhancement}.
Track 1 requires building a single SE system that handles input speech with diverse distortions, domains, and formats. Using pre-defined training datasets, the system must adapt to different sampling frequencies (e.g., 8-48 kHz) and acoustic conditions without prior knowledge of the input type.

Following the previous URGENT challenge\cite{koheiurgent2025}, the task involves seven distortion types, including additive noise, reverberation, clipping, bandwidth limitation, codec distortion, packet loss, and wind noise.
Systems are evaluated on their ability to produce clean, enhanced speech that matches reference signals in quality and intelligibility, or produce high-quality speech in the case of real recordings without reference.
The diversity of speech was further enriched in this challenge, including various emotions, ages, accents, singing voices, whispering speech, and more languages.

\noindent\textbf{Track 2: Speech Quality Assessment}.
Track 2 requires building a speech quality assessment model that predicts MOS for speech enhanced by SE systems. The task focuses on SE-specific distortions and human subjective ratings, rather than synthetic speech or simulated distortions.
Given enhanced speech samples produced by different SE models, systems are evaluated on their ability to accurately predict perceptual speech quality, measured by the correlation with ground-truth MOS annotations both at utterance and system level.

\vspace{-5pt}
\section{Datasets}
\vspace{-5pt}

\textbf{Track 1 Datasets.} Training data is simulated using public speech\footnote{The official dataset of Track 1 is available at \url{https://urgent-challenge.github.io/urgent2026/track1/\#datasets}.}, noise, and room impulse response (RIR) sources, including speech corpora such as LibriVox \cite{kearns2014librivox} (audiobooks), LibriTTS \cite{zenLibriTTSCorpusDerived2019} (reading), Multilingual Librispeech \cite{pratapMLSLargeScaleMultilingual2020} (DE/EN/ES/FR), CommonVoice \cite{ardilaCommonVoiceMassivelyMultilingual2020} (DE/EN/ES/FR/ZH), NNCES (children), VCTK \cite{veauxVoiceBankCorpus2013} (newspaper), EARS \cite{richterEARSAnechoicFullband2024} (studio), SeniorTalk \cite{chenSeniorTalkChineseConversation2025} (elderly), VocalSet \cite{wilkins_2018_1193957} (singing), and ESD \cite{zhouSeenUnseenEmotional2021} (emotional). Noise sources include Audioset \cite{gemmekeAudioSetOntology2017}, WHAM! \cite{wichernWHAMExtendingSpeech2019}, FSD50K~\cite{fonsecaFSD50KOpenDataset2022}, and Free Music Archive~\cite{defferrardFMADatasetMusic2017}. RIRs are simulated from the DNS5 Challenge~\cite{dubeyICASSP2023Deep2024}.

One of the core considerations of Track 1 is to encourage participants to select high-quality data from vast corpora rather than indiscriminately using everything. The baseline curation method~\cite{liLessMoreData2025} selects a 700-hour high-quality speech from the original 2500-hour URGENT 2025 training set~\cite{koheiurgent2025}.

The blind test set contains 360 simulated samples and 480 real-world samples collected from public sources. It includes five unseen languages (HI/KR/AR/JP/IT) beyond those used in the training data.

\noindent\textbf{Track 2 Datasets}.
Training data for Track 2 comprise a diverse collection of MOS-annotated datasets, including synthesized speech (e.g., BC19~\cite{BC19}, SOMOS~\cite{SOMOS}, TTSDS2~\cite{TTSDS2}), voice conversion outputs (BVCC~\cite{BVCC}), speech with simulated or real-world degradations such as telephony artifacts (PSTN~\cite{PSTN}, TCD-VoIP~\cite{TCD-VoIP}) and background noise (TMHINT-QI~\cite{TMHINT-QI}), as well as enhanced speech produced by SE systems (Tencent~\cite{Tencent}, URGENT2024-SQA~\cite{zhangURGENTChallengeUniversality2024, Lessons-Zhang2025}, URGENT2025-SQA~\cite{koheiurgent2025, URGENT-SQA, P808-Sach2025}).
The blind test set contains 8,000 MOS-annotated samples from 16 systems submitted to the URGENT 2025 Challenge blind test phase.

\vspace{-5pt}
\section{Baseline Systems}
\vspace{-5pt}

\textbf{Track 1}. Track 1 baselines\footnote{Both are implemented in PyTorch and available at \url{https://github.com/urgent-challenge/urgent2026_challenge_track1}} include the discriminative BSRNN \cite{yuEfficientMonauralSpeech2023}, equipped with adaptive STFT for sampling-frequency-independent processing, and the generative FlowSE \cite{leeFlowSEFlowMatchingbased2025}, employing conditional flow matching to generate clean speech from noisy inputs.

\noindent\textbf{Track 2}.
The Track 2 baseline\footnote{Pre-trained checkpoints are provided for quick experimentation at \url{https://github.com/urgent-challenge/urgent2026_challenge_track2}} is Uni-VERSA-Ext \cite{uni-versa-ext}, an extension of previous models that incorporates multi-metric supervision for improved MOS prediction. We also report URGENT-PK~\cite{wangURGENTPKPerceptuallyAlignedRanking2025} \footnote{\url{https://github.com/urgent-challenge/URGENT-PK}} in the final leaderboard as an additional baseline just for comparison.

\vspace{-5pt}
\section{Evaluation and Ranking Methods}
\vspace{-5pt}

\textbf{Track 1 Evaluation:} The evaluation follows a two-stage process. In Stage 1 (qualification phase),  we evaluate participants’ submissions with multiple objective metrics using a ranking algorithm inspired by the Friedman test \cite{friedmanUseRanksAvoid1937a}, which is the same as that in the previous URGENT challenges \cite{zhangURGENTChallengeUniversality2024,koheiurgent2025}. The metrics include non-intrusive SE (DNSMOS \cite{reddyDnsmosNonIntrusivePerceptual2021}, NISQA \cite{mittagNISQADeepCNNSelfAttention2021}, UTMOS \cite{saekiUTMOSUTokyoSaruLabSystem2022}, SCOREQ \cite{raganoSCOREQSpeechQuality2024}), intrusive SE (PESQ \cite{ITU-P862}, ESTOI \cite{jensenAlgorithmPredictingIntelligibility2016}, POLQA \cite{ITU-P862}), downstream-task-independent (SpeechBERTScore \cite{saekiSpeechBERTScoreReferenceAwareAutomatic2024a}, LPS~\cite{pirklbauerEvaluationMetricsGenerative2023,P808-Sach2025}), and downstream-task-dependent (Speaker similarity, emotion similarity, language identification accuracy, and character accuracy) ones. 

The top-6 systems from Stage 1 advance to Stage 2 (selection phase), which involves subjective testing using ITU-T P.808 \cite{ITU-P808,P808-Sach2025} ACR and CCR conducted by human listeners on Amazon Mechanical Turk. We rank the top-6 systems based on MOS and CMOS. Statistical significance is assessed to ensure the robustness of the final ranking in-line with earlier ITU-T standardization efforts \cite{ETSI-TS-126-077}.

\noindent\textbf{Track 2 Evaluation:} Systems are evaluated by comparing predicted MOS scores with ground-truth annotations using mean squared error, linear correlation coefficient, Spearman’s rank correlation coefficient, and Kendall’s tau, computed at both utterance and system levels. For final ranking, dense rankings are first computed for each metric, followed by category-wise averaging across error-based and correlation-based metrics to produce the overall system ranking.

\vspace{-5pt}
\section{Results}
\vspace{-5pt}

\textbf{Track 1}.\footnote{Final ranking for both tracks available online at \url{https://urgent-challenge.github.io/urgent2026/ranking/}}  Track 1 received 23 competitive submissions, top six of which progressed to the final subjective evaluation phase. 
Similar to the 2nd URGENT Challenge \cite{koheiurgent2025}, leading systems dominantly followed a hybrid generative and discriminative paradigm. Several top-performing entries employed dual-branch or multi-stage pipelines, leveraging generative models for robust speech restoration and discriminative networks for precise signal preservation. These hybrid frameworks consistently outperformed standalone generative or discriminative baselines across both intrusive and non-intrusive metrics, demonstrating their capability to handle diverse distortions while maintaining speaker and linguistic integrity.

Another key focus was data curation and augmentation. Based on the pre-print technical reports submitted by participants, multiple top-ranked teams emphasized the importance of high-quality training data, utilizing MOS-based filtering, pre-trained restoration models for data cleaning, and tailored augmentation strategies to enhance generalization. This aligns with the challenge’s emphasis on data efficiency and quality-aware training.

\noindent\textbf{Track 2}. 
Track 2 received six competitive submissions, with the top-ranked systems achieving high correlation with human ratings at both system and utterance levels through Uni-VERSA-Ext–style training and improved model components. Notably, the (out-of-competition) URGENT-PK~\cite{wangURGENTPKPerceptuallyAlignedRanking2025} model achieved the highest system-level correlation metrics.

\vspace{-8pt}
\section{Conclusion}
\vspace{-5pt}

The ICASSP 2026 URGENT Speech Enhancement Challenge attracted over 80 teams and received 29 valid submissions across two tracks, emphasizing data curation, speech diversity, and robust evaluation protocols. The results underscore the importance of high-quality training data and hybrid generative-discriminative models for enhancing performance on diverse distortions and languages. Building on previous URGENT challenges, this edition contributes to progress in universal speech enhancement and quality assessment, suggesting directions for future research in scalable and multimodal technologies. Due to space limitations, a more detailed analysis of the results and a more comprehensive overview and analysis of the challenge will be provided in the future.

\vspace{-8pt}

\bibliographystyle{IEEEtran}
\bibliography{refs_v2}

\end{document}